\documentclass[reprint,onecolumn,showpacs,amsmath,amssymb]{revtex4-1}

\usepackage{graphicx}
\usepackage{dcolumn}
\usepackage{bm}
\usepackage{epsfig}
\usepackage{color}
\usepackage{longtable}

\def\etal{et al.}                                        %

\begin{document}

\title{Mass- and field-shift isotope parameters for the $2s  - 2p $ resonance
       doublet of lithium-like ions}

\author{Jiguang Li}
\author{C\'edric Naz\'e}
\author{Michel Godefroid}
\email{mrgodef@ulb.ac.be}
\affiliation{Chimie Quantique et Photophysique, Universit\'e Libre de Bruxelles, B-1050 Brussels, Belgium}

\author{Stephan Fritzsche}
\affiliation{GSI, Helmholtzzentrum f\"ur Schwerionenforschung, D-64291
             Darmstadt, Germany}
\affiliation{FIAS Frankfurt Institute for Advanced Studies,
             D-60438 Frankfurt am Main, Germany}

\author{Gediminas Gaigalas}
\affiliation{Institute of Theoretical Physics and Astronomy, Vilnius University, LT-01108 Vilnius, Lithuania}

\author{Paul Indelicato}
\affiliation{Laboratoire Kastler Brossel, \'Ecole Normale Sup\'erieure, CNRS, and Universit\'e Pierre et Marie Curie-Paris 6, F-75252 Paris CEDEX 05, France}

\author{Per J\"onsson}
\affiliation{School of Technology, Malm\"o University, S-205~06 Malm\"o, Sweden}

\date{\today \\[0.3cm]}

\begin{abstract}
It was recently shown that dielectronic recombination measurements can be used for accurately inferring changes in the nuclear mean-square charge radii of highly-charged lithium-like neodymium [Brandau \etal, \textit{Phys.\ Rev.\ Lett.} 100 073201 (2008)]. To make use of this method to derive information about the nuclear charge distribution for other elements and isotopes, accurate electronic isotope shift parameters are required. In this work, we calculate and discuss the relativistic mass- and field-shift factors for the two $2s \;\, ^2S_{1/2} - 2p \;\, ^2P^o_{1/2,3/2}$ transitions along the lithium isoelectronic sequence. Based on the multiconfiguration Dirac-Hartree-Fock method, the electron correlation and the Breit interaction are taken into account systematically. The analysis of the isotope shifts for these two transitions along the isoelectronic sequence demonstrates the importance and competition between the mass shifts and the field shifts.
\end{abstract}

\maketitle

\section{Introduction} \label{sec:intro}
Isotope shift (IS) measurements have a long tradition and are known as a valuable source for information about (changes in) the nuclear charge radii and distributions. Adding one or several neutrons makes the nuclei not only heavier, but may also modify its charge distribution quite remarkably~\cite{Sanchez:06, Nortershauser:09, Cheal:10, Cocolios:11}. During the last decades, optical laser spectroscopy considerably contributed to the field as an important alternative to a number of earlier x-ray measurements in muonic atoms or electron scattering experiments~\cite{Kluge:10,Cheal2:10}. Using optical laser spectroscopy, measurements became possible for quite long sequences of stable and radioactive isotopes~\cite{Charlwood:10,Avgoulea:11}. A particular merit of this method is that it provides nuclear-model-independent data if the electronic part of the isotope shifts is known sufficiently well. However, the determination of the electronic parameters through ab initio calculations is still a challenge for atomic theory owing to the high sensitivity of the IS parameters with regard to electron correlation and the representation of the wave functions, especially for (nearly-) neutral atoms and ions~\cite{Blondel:01,Carette:11,Carette2:11}.
The electronic contributions to the isotope shifts can be theoretically attained with relatively high precision and reliability for few-electron ions, but as pointed out in Ref. \cite{Angeli:04}, it would be still desirable to have more accurate results for the electronic parameters with up-to-date theoretical methods.

Following the theoretical investigation on the effects of the nuclear charge distribution upon the total cross-section of dielectronic recombination for  Li-like heavy ions  by \c{S}chiopu \etal{}~\cite{Schiopu:04}, isotope shifts of the $2s\,^2S_{1/2} - 2p\,^2P^o_{1/2,3/2}$ transitions have been measured recently at the experimental storage ring at GSI Darmstadt (Germany) for the two stable even-even isotopes $^{142}$Nd and $^{150}$Nd of Li-like neodymium~\cite{Brandau:08}.  From these measurements, it was shown that dielectronic recombination supplies a new technique to derive information on the nuclear charge distributions of heavy stable and unstable elements, provided the electronic parameters are available from theory.
As far as Li-like ions are concerned, many investigations were performed. However, most of them are focused on neutral Li~\cite{Nortershauser:11} (and references therein) and Be$^+$ ion~\cite{Nortershauser:09, Yan:08, Puchalski:08} where nuclear halo structures were predicted some years ago. Godefroid \etal{} have studied the isotope shifts of low-lying levels for Li I through O VI using the multiconfiguration Hartree-Fock method~\cite{Godefroid:01}. Kozhedub \etal{}~\cite{Kozhedub:10} calculated recently the relativistic mass shift for the $2s\,^2S_{1/2} - 2p\,^2P^o_{1/2}$ and $2s\,^2S_{1/2} - 2p\,^2P^o_{3/2}$ transitions along the Li-like isoelectronic sequence. Nevertheless, the isotope shift parameters are scarce for Li-like ions, especially for  the field shift electronic factors.

In this work, we present the electronic parameters relevant to the isotope shift for the two $2s \;\, ^2S_{1/2} - 2p \;\, ^2P^o_{1/2,3/2}$ transitions in Li-like ions in the range from $Z=4$ to $Z=90$. The electron correlation and Breit interaction are treated within the framework of the multiconfiguration Dirac-Hartree-Fock (MCDF) method. Comparison is made with other theoretical predictions, when available. Using our calculated  parameters the balance between the mass shifts and the field shifts of these two transitions is discussed along the isoelectronic sequence with the assistance of the simple empirical formulae for the nuclear mass number and the nuclear mean-square radius. In addition we investigate in great detail the case of $^{150,142}$Nd$^{57+}$ for which experimental IS measurements were recently reported~\cite{Brandau:08}.

\section{Theory and computational strategy} \label{sec:theo}
\subsection{Isotope Shift Theory}
In the approximation of  the first-order perturbation theory, the isotope shift, for a transition involving the atomic states $u$ and $l$ between two isotopes with mass numbers $A > A'$, can be parameterized as~\cite{Fricke:95}
\begin{equation}
\label{MF-eq}
   \delta \nu^{A,A'}  =  \Delta \tilde{K}^{RMS} \; \frac{M'-M}{MM'} \;+\; F \delta \langle r^2 \rangle^{A,A'} \; ,
\end{equation}
where $\Delta \tilde{K}^{RMS} \,=\, (\tilde{K}^{RMS}_u - \tilde{K}^{RMS}_l) $ refers to the relativistic mass shift parameter and $F$ is the field shift factor; $M'$ and $M$ are the nuclear masses and $\delta \langle r^2 \rangle^{A,A'}$ represents the nuclear mean-square radius difference between two isotopes $A$ and $A'$.

The mass-shift parameter $\tilde{K} $ is related to the expectation value of the relativistic recoil Hamiltonian
\begin{equation}
\tilde{K}^{RMS} = \frac{M}{h} \langle \Psi |  H_{RMS} | \Psi \rangle   \; .
\end{equation}
The recoil Hamiltonian, correct to the order of $(m/M)$ in the mass ratio of the electron and nucleus, is given by~\cite{Palmer:87, Shabaev:94, Shabaev:98, Gaidamauskas:11}
\begin{eqnarray}\label{RNRO}
H_{RMS} = \frac{1}{2M} \sum_{i,j}\left\{{\bm p}_i \cdot {\bm p}_j - \frac{\alpha Z}{r_i} \left[{\bm \alpha}_i + \frac{({\bm \alpha}_i \cdot {\bm r}_i){\bm r}_i}{r_i^2} \right] \cdot {\bm p}_j \right\} \; ,
\end{eqnarray}
and is split into the one-body and the two-body relativistic mass shift operators
\begin{eqnarray}
\label{RNMS} H_{RNMS} &=& \frac{1}{2M} \sum_{i} \left\{ {\bm p}_i^2 - \frac{\alpha Z}{r_i} \left[ {\bm \alpha}_i + \frac{({\bm \alpha}_i \cdot {\bm r}_i) {\bm r}_i}{r_i^2} \right] \cdot {\bm p}_i \right\}, \\
\label{RSMS} H_{RSMS} &=& \frac{1}{2M} \sum_{i \neq j} \left\{  {\bm p}_i \cdot {\bm p}_j - \frac{\alpha Z}{r_i} \left[ {\bm \alpha}_i + \frac{({\bm \alpha}_i \cdot {\bm r}_i){\bm r}_i}{r_i^2} \right] \cdot {\bm p}_j \right\} \; ,
\end{eqnarray}
referred respectively as the relativistic normal mass shift (RNMS) and specific mass shift (RSMS) terms.
Adopting this partition, the transition isotope shift (\ref{MF-eq}) becomes
\begin{equation}
\label{MF-eq_2}
   \delta \nu^{A,A'}  = (\Delta \tilde{K}^{RNMS} + \Delta \tilde{K}^{RSMS}) \, \frac{M'-M}{MM'} \;+\; F \delta \langle r^2 \rangle^{A,A'} \; .
\end{equation}

Note that the normal mass shift values reported in the present work are estimated using the perturbation approach from the expectation values of the RNMS operator rather than adopting  the scaling law ($ \Delta \tilde{K}^{NMS} = m_e \nu$)
that becomes inappropriate for heavy relativistic systems~\cite{Palmer:87, Veitia:04, Li:12}.

The field shift factor appearing in Eq.~(\ref{MF-eq}) has the form~\cite{Torbohm:85}
\begin{eqnarray}
F = \frac{Z}{3 \hbar}  \left( \frac{e^2}{4 \pi \epsilon_0} \right) \Delta|\Psi({\bm 0})|^2 \; ,
\end{eqnarray}
 which is proportional to the change of the electronic total probability density $\rho^e(\bm 0)$ at the origin between the two atomic states involved in the transition, i.e.~$\Delta |\Psi({\bm 0})|^2 = \rho^e_u({\bm 0}) - \rho^e_l({\bm 0})$,
where $\rho^e ({\bm 0})$ is defined by
\begin{eqnarray}
\rho^e ({\bm 0}) = \langle \Psi |\sum_{i}\delta({\bm r}_i)| \Psi \rangle.
\end{eqnarray}

\subsection{MCDF Method}
The multiconfiguration Dirac-Hartree-Fock (MCDF) method~\cite{Grant:book} is employed to yield atomic state wave functions (ASFs). On the basis of the MCDF method, the ASF is represented by a linear combination of configuration state functions (CSFs) with same parity $P$, total angular momentum $J$ and its component along $z$-direction $M_J$ as
\begin{equation}
\Psi(P J M_J) = \sum_{i=1}^{N_c} c_i \Phi(\gamma_i P J M_J),
\end{equation}
where $\{ c_i \} $ are the mixing coefficients and $\{ \gamma_i \}$  the sets of other quantum numbers needed for specifying the $N_c$ CSFs. The latter are built from single-electron orbital wave functions. Applying the variational principle, the mixing coefficients $c_i$ and single-electron orbital wave functions are optimized simultaneously via the self-consistent field (SCF) method. A more limited version is the relativistic configuration interaction (RCI) approach allowing only the mixing coefficients to be varied. In the RCI computation, the Breit interaction can be taken into account perturbatively as well.

It should be emphasized that instead of a point-like nucleus, a finite nuclear charge distribution model, for instance, a uniform spherical model or a two-parameters Fermi model, must be used to generate the nuclear potential in the calculation in order to ensure the first-order perturbation to be valid for the field shift~\cite{Blundell:87}. In the present work, the two-parameters Fermi nuclear model
\begin{eqnarray}\label{TPFNM}
\rho (r) = \frac{\rho_0}{1 + e^{(r-c)/a}}
\end{eqnarray}
is adopted. Here $\rho_0$ is a normalization coefficient, $c$ is the half-density radius and $a = \frac{t}{4 \ln 3}$ is related to the surface thickness $t$ of the charge distribution. In practice, the  $t = 2.30\,\mathrm{fm}$ value is used and $c$ is computed according to the formulae given in Ref. \cite{Parpia:92}. Since the ASFs are insensitive with respect to details of the nuclear model \cite{Blundell:87, Shabaev:93}, the nuclear parameters from any stable isotope can be chosen to perform the SCF and RCI calculations.

\subsection{Computational Models} \label{ssec:cm}

The active space  approach is utilized for building the ASF in order to capture the main correlation effects systematically. In this framework, the configuration space is expanded by means of single (S) and double (D) excitations from occupied orbitals to some active set. The latter is augmented layer by layer, which makes it possible to monitor the convergence of the physical quantity concerned. Each correlation layer, labelled by $n=2,3,4, \ldots$, contains the $s,\, p,\, d,\, \ldots$ orbitals ($l_{\mbox{max}}=6$). Starting with Dirac-Hartree-Fock (DF) calculations, MCDF calculations are carried out sequentially to optimize the correlation orbitals in the added layer, together with the mixing coefficients. The Breit interaction is taken into account in the subsequent RCI computations. The calculations are performed by using the GRASP2K package~\cite{GRASP2K,GRASP2K:12} in which relativistic mass shifts corrections are implemented~\cite{Gaidamauskas:11,Naze:12}. For cross-checking our investigations on some cases, we also use the MCDF-GME package~\cite{Indelicato:MCDF}.

\section{Results and Discussion} \label{sec:results}
\subsection{Electron correlation effects on isotope shifts} \label{CE}
Being interested in the entire isoelectronic lithium sequence and realizing the importance of correlation effects in the neutral end, we test our computational strategy on Be$^+$. For this ion, we present in table \ref{tab:1} the relativistic normal mass shift ($\Delta \tilde{K}^{RNMS}$), the relativistic specific mass shift ($\Delta \tilde{K}^{RSMS}$) and the field shift ($F$) factors for the $2s \;\, ^2S_{1/2} - 2p \;\, ^2P^o_{1/2}$ and $2s \;\, ^2S_{1/2} - 2p \;\, ^2P^o_{3/2}$ transitions, omitting the Breit interaction in the calculations. As can be seen from this table, SD excitations up to $n=8$ are essential to bring these values to the demanded precision. However, one observes that the relativistic normal mass shift in these transitions is more sensitive to correlation due to large cancellation when making the difference between the level shifts considered. Complete convergence of these physical quantities is considerably hard to achieve, but we control the uncertainties within $\approx 1$~\% for the total mass shift and the field shift.

Furthermore, we perform additional RCI calculations by appending the triple (T) excitations from the occupied orbitals to the active set of $n=5$, to the largest SD configuration space ($n=8$). The corresponding values marked as SDT are displayed in the same table. It is found that the contribution of triple excitations is definitely fractional. We should keep in mind that some orbitals are kept frozen in the layer-by-layer calculations to avoid convergence problems. The corresponding loss of orbital relaxation can be estimated with the MCDF-GME package~\cite{Indelicato:MCDF} that allows to optimize all orbitals variationally. In this calculation, the SDT excitations up to $n=8$ are used to generate the configuration space. The MCDF-GME results are listed in the last row in table~\ref{tab:1}. The satisfactory consistency between the results obtained by these two independent codes is on line with the 1\% estimated uncertainty for the calculated shifts and demonstrates the adequacy of the computation strategy described in Section~\ref{ssec:cm} that is used for heavier ions.

\begin{table}
\caption{Relativistic normal mass shift $\Delta \tilde{K}^{RNMS}$ (in~GHz~u), specific mass shift $\Delta \tilde{K}^{RSMS}$  (in~GHz~u) and field shift $F$ (in MHz/fm$^2$) factors for the $2s \;\, ^2S_{1/2} - 2p \;\, ^2P^o_{1/2}$ and $2s \;\, ^2S_{1/2} - 2p \;\, ^2P^o_{3/2}$ transitions in  Be$^+$. The label $(n=x)$ specifies the orbital active set.}
\begin{center}
\begin{tabular}{l l  c  c c c c c  c   c c c c c} \\
\hline \hline \\[-0.2cm]
                                    &&& \multicolumn{3}{c}{$2s \;\, ^2S_{1/2} - 2p \;\, ^2P^o_{1/2}$}
                                     && \multicolumn{3}{c}{$2s \;\, ^2S_{1/2} - 2p \;\, ^2P^o_{3/2}$} \\[0.1cm]
\cline{4-6} \cline{8-10}  \\[-0.2cm]
Code     &       Model               &&    $\Delta \tilde{K}^{RNMS}$  &  $\Delta \tilde{K}^{RSMS}$    &   $F$
                                     &\hspace*{0.3cm} &    $\Delta \tilde{K}^{RNMS}$     &   $\Delta \tilde{K}^{RSMS}$  &   $F$        \\[0.1cm]
\hline  \\[-0.2cm]
         &        DF                 && $-529.1$ &  $-957.9$  &   $-16.72$
                                     && $-529.1$ &  $-958.0$  &   $-16.72$   \\
         &      SD(n=2)              &&   $-521$ &  $-2610 $  &   $-15.39$
                                     &&   $-521$ &  $-2610 $  &   $-15.39$   \\
GRASP2K  &      SD(n=7)              &&   $-529$ &  $-1028 $  &   $-16.94$
                                     &&   $-529$ &  $-1028 $  &   $-16.94$   \\
         &      SD(n=8)              &&   $-538$ &  $-1025 $  &   $-16.98$
                                     &&   $-538$ &  $-1025 $  &   $-16.98$   \\
         &   $\bigcup$ SDT(n=5)      &&   $-538$ &  $-1025 $  &   $-16.99$
                                     &&   $-538$ &  $-1025 $  &   $-16.99$   \\[0.1cm]
\hline \\[0.01cm]
MCDF-GME & SDT(n=8)                  &&   $-532$ &  $-1034$   &   $-17.05$
                                     &&   $-534$ &  $-1034$   &   $-17.15$   \\[0.1cm]
\hline \hline \\
\end{tabular}
\end{center}
\label{tab:1}
\end{table}

\subsection{Electronic Factors in Isotope Shifts}
In table \ref{tab:2} we present the total mass shift $\Delta \tilde{K}^{RMS}$ and field shift $F$ factors for the $2s \;\, ^2S_{1/2} - 2p \;\, ^2P^o_{1/2}$ and $2s \;\, ^2S_{1/2} - 2p \;\, ^2P^o_{3/2}$ in Be$^{+}$, Zn$^{27+}$, Nd$^{57+}$ and Hg$^{77+}$. In addition, the individual values of relativistic normal mass shifts $\Delta \tilde{K}^{RNMS}$ and specific mass shifts $\Delta \tilde{K}^{RSMS}$ are reported. To illustrate the effects of electron correlation and Breit interaction on these physical quantities along the isoelectronic sequence, the calculations are carried out using  different computational models, namely, DF, MCDF, and MCDF+Breit. The differences between the MCDF and DF results reflect the correlations between the electrons. Since the correlation effects become usually less important for high-$Z$ ions, we have reduced gradually the size of active orbital set with increasing the nuclear charge. Based on the MCDF model, the Breit interaction is included into the values denoted by the ``MCDF+Breit'' label. It is observed that both the electron correlation and the Breit interaction contribute little to the total mass shift and field shift factors in these two transitions, although the effect is larger for the  $\Delta \tilde{K}^{RNMS}$ contribution. Furthermore, we find out that the two-body part of the mass shift operator is at least two orders of magnitude larger than the one-body part in the  middle- and high-$Z$ region. This is in opposition with Seltzer~\cite{Seltzer:69} who pointed out that, in the case of $K$ x-ray transitions, the specific mass shift is approximately $- 1/3$ of the normal mass shift for $40 < Z < 70$.

We also compare our results with other available theoretical calculations \cite{Godefroid:01, Yan:08, Puchalski:08, Kozhedub:10, Brandau:08} that are collected in table~\ref{tab:2}. As can be seen from this table, the agreement is satisfactory for the total mass shift $\Delta \tilde{K}^{RMS}$ and the field shift $F$ factors. However, we notice a quite large discrepancy with Kozhedub \etal{}'s value \cite{Kozhedub:10} in the RNMS of the $2s \;\, ^2S_{1/2} - 2p \;\, ^2P^o_{1/2}$ transition in Nd$^{57+}$. Additionally,
the differences between our Nd$^{57+}$ field-shift factors and Brandau \etal{} results~\cite{Brandau:08}  are relatively small (about 5\%). Independent calculations \footnote{In the calculations, the projector operator has been used to avoid the convergence problems encountered in the SCF procedure [P. Indelicato, Phys. Rev. A 51, 1132 (1995)].}, carried out using the MCDF-GME package, are consistent with GRASP2K.


The total results, obtained by the three computational models, for the relativistic mass shift and the field shift factors of the $2s \;\, ^2S_{1/2} - 2p \;\, ^2P^o_{1/2}$ and $2s \;\, ^2S_{1/2} - 2p \;\, ^2P^o_{3/2}$ transitions are presented in table \ref{tab:3} and \ref{tab:4}
from $Z=4$ to $Z=90$. It is worth noting that the trend of $\Delta \tilde{K}^{RMS}$ for the $2s \;\, ^2S_{1/2} - 2p \;\, ^2P^o_{1/2}$ transition is not monotonous due to the fact that the relativistic corrections to the mass shift become important in the heavy elements. For this line, the contribution of the second term of Eq. (\ref{RNRO}) to the total line mass shift, counteracting the first term, increases drastically with the nuclear charge (10\% in Fe$^{23+}$, 47\% in Nd$^{57+}$ and 90\% in Th$^{87+}$). Note that the QED contribution to the recoil operator, calculated by Kozhedub \etal{}~\cite{Kozhedub:10} and omitted in the present work, becomes larger than the relativistic contribution for the high-$Z$ region. The field shift factors for these two transitions however increase considerably rapidly along the isoelectronic sequence.

The uncertainties of the mass- and field-shift parameters presented in table \ref{tab:3} and \ref{tab:4} mainly result from the electron correlation effects for the low-$Z$ ions. As discussed in section \ref{CE}, the effects lead to about 1\% errors along the isoelectronic sequence. Moreover, we neglect the frequency-dependent Breit interaction (FDBI) and the quantum electrodynamical (QED) corrections in the present calculations and these increase for the high-$Z$ ions. We roughly estimate their contributions to the factors for Li-like Nd and Th ions. It is found that the FDBI and QED corrections affect the field shift factors by 1\% for Li-like Nd and 3\% for Li-like Th in both transitions under consideration. Also, they give 1\% and 5\% contributions to the mass shift factor in the $2s \;\, ^2S_{1/2} - 2p \;\, ^2P^o_{1/2}$ transition and 1\% and 3\% in the $2s \;\, ^2S_{1/2} - 2p \;\, ^2P^o_{3/2}$ transition for Li-like Nd and Th ions, respectively.


\begin{table}[!ht]
\begin{scriptsize}
\caption{Relativistic mass shift $\Delta \tilde{K}^{RMS}$  (in GHz u) and field shift  $F$ (in MHz/fm$^2$) factors for the $2s \;\, ^2S_{1/2} - 2p \;\, ^2P^o_{1/2}$ and $2s \;\, ^2S_{1/2} - 2p \;\, ^2P^o_{3/2}$ in Be$^{+}$, Zn$^{27+}$, Nd$^{57+}$ and Hg$^{77+}$. For comparison, individual relativistic normal mass shift $\Delta \tilde{K}^{RNMS}$ and specific mass shift coefficient $\Delta \tilde{K}^{RSMS}$ (in GHz u) are presented as well. The label of $n$ in parentheses represents the largest size of the active set. The number in square brackets is the power of 10.}
\begin{center}
\begin{tabular}{l  c  c c c c c  c   c c c c c} \\
\hline \hline \\[-0.2cm]
Model                               && \multicolumn{4}{c}{$2s \;\, ^2S_{1/2} - 2p \;\, ^2P^o_{1/2}$}
                                     && \multicolumn{4}{c}{$2s \;\, ^2S_{1/2} - 2p \;\, ^2P^o_{3/2}$} \\[0.1cm]
\cline{3-6} \cline{8-11}  \\[-0.2cm]
                                     &&    $\Delta \tilde{K}^{RNMS}$  &  $\Delta \tilde{K}^{RSMS}$    & $\Delta \tilde{K}^{RMS}$ &   $F$
                                     &\hspace*{0.3cm} &    $\Delta \tilde{K}^{RNMS}$     &   $\Delta \tilde{K}^{RSMS}$  & $\Delta \tilde{K}^{RMS}$ &  $F$  \\[0.1cm]
\hline  \\[-0.2cm]
\multicolumn{11}{c}{\textbf{Be}$^{\bm +}$} \\
DF                                   &&   $-529.1$ &  $-957.9$  & $-1487$ & $-16.72$
                                     &&   $-529.1$ &  $-958.0$  & $-1487$ & $-16.72$       \\
MCDF(n=8)                            &&   $-537.8$ &  $-1025 $  & $-1562$ & $-16.98$
                                     &&   $-538.0$ &  $-1025 $  & $-1563$ & $-16.98$       \\
MCDF+Breit                           &&   $-537.8$ &  $-1025 $  & $-1562$ & $-16.97$
                                     &&   $-537.9$ &  $-1025 $  & $-1563$ & $-16.97$       \\
Godefroid \etal{} \cite{Godefroid:01}&&            &  $-1035 $  &         & $-16.91$
                                     &&            &  $-1035 $  &         & $-16.91$       \\
Yan \etal{}  \cite{Yan:08}           &&            &            &         & $-16.912$
                                     &&            &            &         & $-16.912$      \\[0.1cm]
\hline \\[-0.2cm]
\multicolumn{11}{c}{\textbf{Zn}$^{\bm{27+}}$} \\
DF                                   &&  $-7475      $ & $-2177[2]    $ & $-2251[2]      $ & $-2009[2]$
                                     &&  $-1143[1]   $ & $-2194[2]    $ & $-2308[2]      $ & $-2025[2]$   \\
MCDF(n=6)                            &&  $-7727      $ & $-2168[2]    $ & $-2245[2]      $ & $-2012[2]$
                                     &&  $-1128[1]   $ & $-2187[2]    $ & $-2300[2]      $ & $-2024[2]$   \\
MCDF+Breit                           &&  $-7895      $ & $-2167[2]    $ & $-2246[2]      $ & $-2009[2]$
                                     &&  $-1129[1]   $ & $-2188[2]    $ & $-2301[2]      $ & $-2021[2]$   \\
Kozhedub \etal{}  \cite{Kozhedub:10} &&  $-8054.2    $ & $-2165.449[2]$ & $-2246.00(3)[2]$ &
                                     &&  $-1124.97[1]$ & $-2168.232[2]$ & $-2300.73(3)[2]$ &            \\[0.1cm]
\hline \\[-0.2cm]
\multicolumn{11}{c}{\textbf{Nd}$^{\bm{57+}}$} \\
DF                                   &&  $-1083[1]  $ &  $-8227[2]   $ & $-8336[2]       $ &  $-7903[3]$
                                     &&  $-8721[1]  $ &  $-8768[2]   $ & $-9640[2]       $ &  $-8215[3]$   \\
MCDF(n=5)                            &&  $-1342[1]  $ &  $-8196[2]   $ & $-8331[2]       $ &  $-7929[3]$
                                     &&  $-8589[1]  $ &  $-8761[2]   $ & $-9620[2]       $ &  $-8203[3]$   \\
MCDF+Breit                           &&  $-1449[1]  $ &  $-8196[2]   $ & $-8341[2]       $ &  $-7885[3]$
                                     &&  $-8577[1]  $ &  $-8775[2]   $ & $-9632[2]       $ &  $-8157[3]$   \\
Kozhedub \etal{} \cite{Kozhedub:10}  &&  $-1641.8[1]$ &  $-8180.90[2]$ & $-8345.08(25)[2]$ &
                                     &&  $-8573.3[1]$ &  $-8769.29[2]$ & $-9626.62(25)[2]$ &             \\
Brandau \etal{} \cite{Brandau:08}    &&               &                &                   &  $-7520[3]$
                                     &&               &                &                   &  $-7810[3]$   \\
MCDF-GME                             &&               &                &                   &  $-7897[3]$
                                     &&               &                &                   &  $-8168[3]$   \\[0.1cm]
\hline \\[-0.2cm]
\multicolumn{11}{c}{\textbf{Hg}$^{\bm{77+}}$} \\
DF                                   && $  4691[1]$   &  $-1146[3]$  & $-1099[3]  $ &  $-5761[4]$
                                     && $ -2193[2]$   &  $-1452[3]$  & $-1671[3]  $ &  $-6239[4]$   \\
MCDF(n=4)                            && $  3990[1]$   &  $-1140[3]$  & $-1100[3]  $ &  $-5791[4]$
                                     && $ -2158[2]$   &  $-1451[3]$  & $-1667[3]  $ &  $-6225[4]$   \\
MCDF+Breit                           && $  2961[1]$   &  $-1119[3]$  & $-1090[3]  $ &  $-5783[4]$
                                     && $ -2181[2]$   &  $-1434[3]$  & $-1652[3]  $ &  $-6213[4]$   \\
Kozhedub \etal{} \cite{Kozhedub:10}  &&               &              & $-1105.6[3]$ &
                                     &&               &              & $-1669.5[3]$ &               \\[0.1cm]
\hline
\hline
\end{tabular}
\end{center}
\label{tab:2}
\end{scriptsize}
\end{table}

\begin{table}[!ht]
\caption{Mass- and field-shift factors, $\Delta \tilde{K}^{RMS}$ (in GHz u) and $F$ (in MHz/fm$^2$), for the $2s \;\, ^2S_{1/2} - 2p \;\, ^2P^o_{1/2}$ transition for selected elements along the lithium isoelectronic sequence. The number in the square brackets represents the power of 10.}
\begin{center}
\begin{tabular}{l c c c c c c c c} \\
\hline \hline \\[-0.2cm]
           && \multicolumn{3}{c}{$\Delta \tilde{K}^{RMS}$} &\hspace*{0.4cm} &  \multicolumn{3}{c}{$F$} \\[0.1cm]
\cline{3-5} \cline{7-9} \\[-0.2cm]
           &&    DF         &   MCDF        &   MCDF+Breit  &&     DF       &   MCDF         &   MCDF+Breit       \\[0.1cm]
\hline  \\[-0.2cm]
Be$^{+}$   && $-1.487[3]$  &  $-1.562[3]$  &  $-1.562[3]$  && $-1.672[1]$  &  $-1.698[1]$   &   $-1.697[1]$      \\
C$^{3+}$   && $-5.540[3]$  &  $-5.581[3]$  &  $-5.581[3]$  && $-1.408[2]$  &  $-1.418[2]$   &   $-1.417[2]$      \\
O$^{5+}$   && $-1.190[4]$  &  $-1.188[4]$  &  $-1.188[4]$  && $-5.434[2]$  &  $-5.454[2]$   &   $-5.451[2]$      \\
Ne$^{7+}$  && $-2.050[4]$  &  $-2.042[4]$  &  $-2.042[4]$  && $-1.486[3]$  &  $-1.489[3]$   &   $-1.488[3]$      \\
Si$^{11+}$ && $-4.438[4]$  &  $-4.417[4]$  &  $-4.416[4]$  && $-6.566[3]$  &  $-6.575[3]$   &   $-6.571[3]$      \\
Ar$^{15+}$ && $-7.701[4]$  &  $-7.668[4]$  &  $-7.669[4]$  && $-1.983[4]$  &  $-1.986[4]$   &   $-1.984[4]$      \\
Ti$^{19+}$ && $-1.182[5]$  &  $-1.178[5]$  &  $-1.178[5]$  && $-4.844[4]$  &  $-4.850[4]$   &   $-4.844[4]$      \\
Fe$^{23+}$ && $-1.677[5]$  &  $-1.672[5]$  &  $-1.672[5]$  && $-1.033[5]$  &  $-1.035[5]$   &   $-1.033[5]$      \\
Zn$^{27+}$ && $-2.251[5]$  &  $-2.245[5]$  &  $-2.246[5]$  && $-2.009[5]$  &  $-2.012[5]$   &   $-2.009[5]$      \\
Kr$^{33+}$ && $-3.252[5]$  &  $-3.244[5]$  &  $-3.245[5]$  && $-4.824[5]$  &  $-4.833[5]$   &   $-4.822[5]$      \\
Mo$^{39+}$ && $-4.399[5]$  &  $-4.392[5]$  &  $-4.393[5]$  && $-1.052[6]$  &  $-1.054[6]$   &   $-1.051[6]$      \\
Xe$^{51+}$ && $-6.997[5]$  &  $-6.991[5]$  &  $-6.988[5]$  && $-4.176[6]$  &  $-4.188[6]$   &   $-4.167[6]$      \\
Nd$^{57+}$ && $-8.336[5]$  &  $-8.331[5]$  &  $-8.341[5]$  && $-7.903[6]$  &  $-7.929[6]$   &   $-7.885[6]$      \\
Yb$^{67+}$ && $-1.027[6]$  &  $-1.027[6]$  &  $-1.028[6]$  && $-2.165[7]$  &  $-2.174[7]$   &   $-2.172[7]$      \\
Hg$^{77+}$ && $-1.099[6]$  &  $-1.100[6]$  &  $-1.090[6]$  && $-5.761[7]$  &  $-5.791[7]$   &   $-5.783[7]$      \\
Bi$^{80+}$ && $-1.072[6]$  &  $-1.073[6]$  &  $-1.053[6]$  && $-7.711[7]$  &  $-7.752[7]$   &   $-7.741[7]$      \\
Fr$^{84+}$ && $-9.772[5]$  &  $-9.779[5]$  &  $-9.379[5]$  && $-1.133[8]$  &  $-1.139[8]$   &   $-1.137[8]$      \\
Th$^{87+}$ && $-8.465[5]$  &  $-8.472[5]$  &  $-7.832[5]$  && $-1.512[8]$  &  $-1.521[8]$   &   $-1.518[8]$      \\[0.1cm]
\hline \hline \\
\end{tabular}
\end{center}
\label{tab:3}
\end{table}

\begin{table}[!ht]
\caption{Mass- and field-shift factors, $\Delta \tilde{K}^{RMS}$ (in GHz u) and $F$ (in MHz/fm$^2$), for the $2s \;\, ^2S_{1/2} - 2p \;\, ^2P^o_{3/2}$ transition for selected elements along the lithium isoelectronic sequence. The number in the square brackets represents the power of 10.}
\begin{center}
\begin{tabular}{l c c c c c c c c} \\
\hline \hline \\[-0.2cm]
           && \multicolumn{3}{c}{$\Delta \tilde{K}^{RMS}$} &\hspace*{0.4cm} &  \multicolumn{3}{c}{$F$} \\[0.1cm]
\cline{3-5} \cline{7-9} \\[-0.2cm]
           &&    DF        &    MCDF       &  MCDF+Breit   &&   DF         &     MCDF       &   MCDF+Breit       \\[0.1cm]
\hline  \\[-0.2cm]
Be$^{+}$   && $-1.487[3]$  &  $-1.563[3]$  &  $-1.563[3]$  && $-1.672[1]$  &  $-1.698[1]$   &   $-1.697[1]$      \\
C$^{3+}$   && $-5.543[3]$  &  $-5.583[3]$  &  $-5.584[3]$  && $-1.409[2]$  &  $-1.418[2]$   &   $-1.417[2]$      \\
O$^{5+}$   && $-1.191[4]$  &  $-1.189[4]$  &  $-1.189[4]$  && $-5.436[2]$  &  $-5.453[2]$   &   $-5.451[2]$      \\
Ne$^{7+}$  && $-2.054[4]$  &  $-2.046[4]$  &  $-2.046[4]$  && $-1.487[3]$  &  $-1.489[3]$   &   $-1.489[3]$      \\
Si$^{11+}$ && $-4.457[4]$  &  $-4.435[4]$  &  $-4.434[4]$  && $-6.576[3]$  &  $-6.579[3]$   &   $-6.575[3]$      \\
Ar$^{15+}$ && $-7.761[4]$  &  $-7.726[4]$  &  $-7.727[4]$  && $-1.989[4]$  &  $-1.989[4]$   &   $-1.987[4]$      \\
Ti$^{19+}$ && $-1.197[5]$  &  $-1.192[5]$  &  $-1.192[5]$  && $-4.864[4]$  &  $-4.863[4]$   &   $-4.857[4]$      \\
Fe$^{23+}$ && $-1.707[5]$  &  $-1.701[5]$  &  $-1.701[5]$  && $-1.039[5]$  &  $-1.039[5]$   &   $-1.037[5]$      \\
Zn$^{27+}$ && $-2.308[5]$  &  $-2.300[5]$  &  $-2.301[5]$  && $-2.025[5]$  &  $-2.024[5]$   &   $-2.021[5]$      \\
Kr$^{33+}$ && $-3.378[5]$  &  $-3.368[5]$  &  $-3.369[5]$  && $-4.883[5]$  &  $-4.880[5]$   &   $-4.869[5]$      \\
Mo$^{39+}$ && $-4.649[5]$  &  $-4.638[5]$  &  $-4.640[5]$  && $-1.070[6]$  &  $-1.069[6]$   &   $-1.065[6]$      \\
Xe$^{51+}$ && $-7.787[5]$  &  $-7.771[5]$  &  $-7.771[5]$  && $-4.304[6]$  &  $-4.298[6]$   &   $-4.278[6]$      \\
Nd$^{57+}$ && $-9.640[5]$  &  $-9.620[5]$  &  $-9.632[5]$  && $-8.215[6]$  &  $-8.203[6]$   &   $-8.157[6]$      \\
Yb$^{67+}$ && $-1.307[6]$  &  $-1.305[6]$  &  $-1.306[6]$  && $-2.291[7]$  &  $-2.286[7]$   &   $-2.283[7]$      \\
Hg$^{77+}$ && $-1.671[6]$  &  $-1.667[6]$  &  $-1.652[6]$  && $-6.239[7]$  &  $-6.225[7]$   &   $-6.213[7]$      \\
Bi$^{80+}$ && $-1.775[6]$  &  $-1.770[6]$  &  $-1.745[6]$  && $-8.422[7]$  &  $-8.401[7]$   &   $-8.383[7]$      \\
Fr$^{84+}$ && $-1.900[6]$  &  $-1.894[6]$  &  $-1.846[6]$  && $-1.253[8]$  &  $-1.250[8]$   &   $-1.247[8]$      \\
Th$^{87+}$ && $-1.977[6]$  &  $-1.970[6]$  &  $-1.896[6]$  && $-1.690[8]$  &  $-1.685[8]$   &   $-1.681[8]$      \\[0.1cm]
\hline \hline \\
\end{tabular}
\end{center}
\label{tab:4}
\end{table}

%
%
%
%

\subsection{Mass and Field Shifts Balance}\label{MF}
The values of the mass shift and the field shift depend not only on the specific isotope pair, but on the transition under investigation as well. Consequently, the total isotope shift relies on the relative sign of the mass and field shifts if they are of the similar order of magnitude. In fact, the mass shift and the field shift are subject to different trends along the isoelectronic sequence, which makes it important to discuss the competition between these two expectation values in different $Z$-regions.

To evaluate the mass shifts and the field shifts for the $2s \;\, ^2S_{1/2} - 2p \;\, ^2P^o_{1/2}$ and $2s \;\, ^2S_{1/2} - 2p \;\, ^2P^o_{3/2}$ transitions along the Li-like isoelectronic sequence, the difference of the nuclear mass and mean-square charge radius for a given isotope pair should be determined. For convenience, we use the mass number $A$ instead of the nuclear mass and adopt the empirical formula \cite{Schiopu:04}
\begin{eqnarray}\label{AEF}
Z = \frac{A}{1.98+0.015A^{2/3}}.
\end{eqnarray}
to inversely deduce $A$ for a stable isotope of given atom with the atomic number $Z$. Considering two isotopes of the selected element, the $\delta \langle r^2 \rangle^{A,A'}$ can be obtained via the empirical formula (in fm)~\cite{Andrae:00}
\begin{eqnarray}\label{NREF}
\langle r^2 \rangle^{1/2}  = 0.836 A^{1/3} + 0.570\ (A > 9) \; .
\end{eqnarray}

\begin{figure}[!ht]
\vspace*{1.0cm}

\includegraphics[scale=0.8]{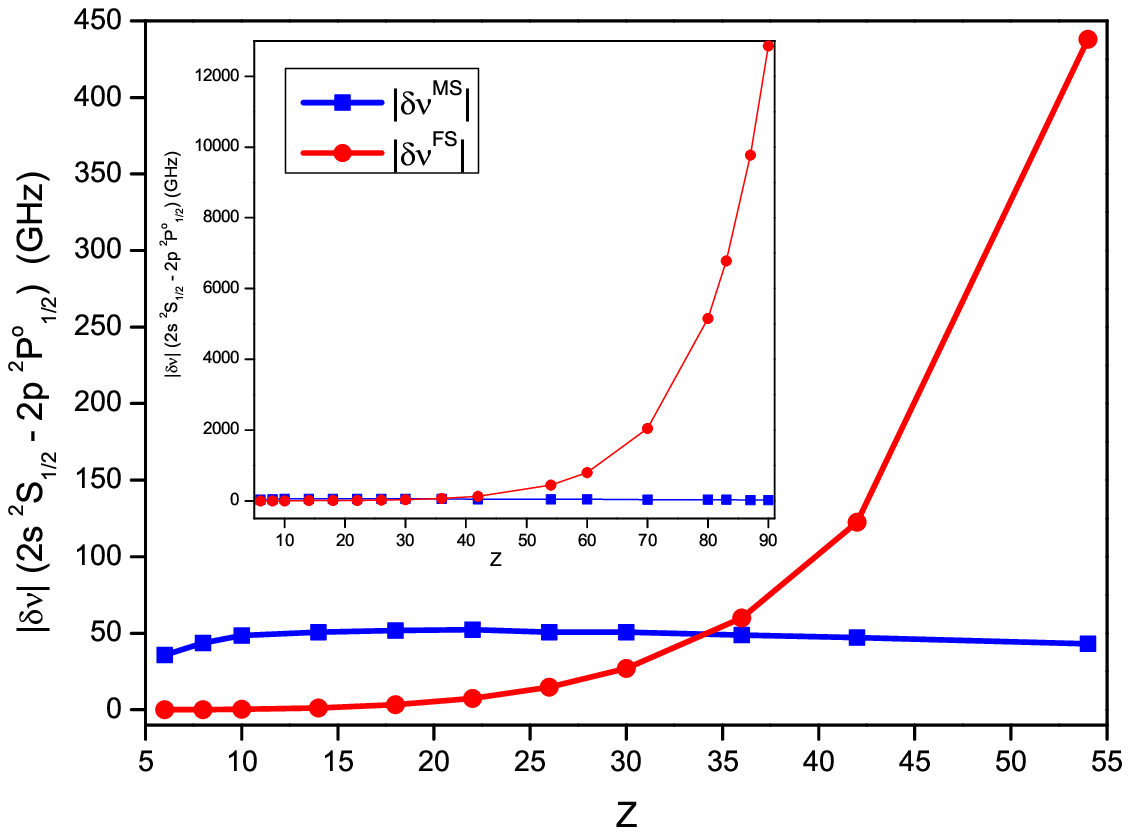}


\includegraphics[scale=0.8]{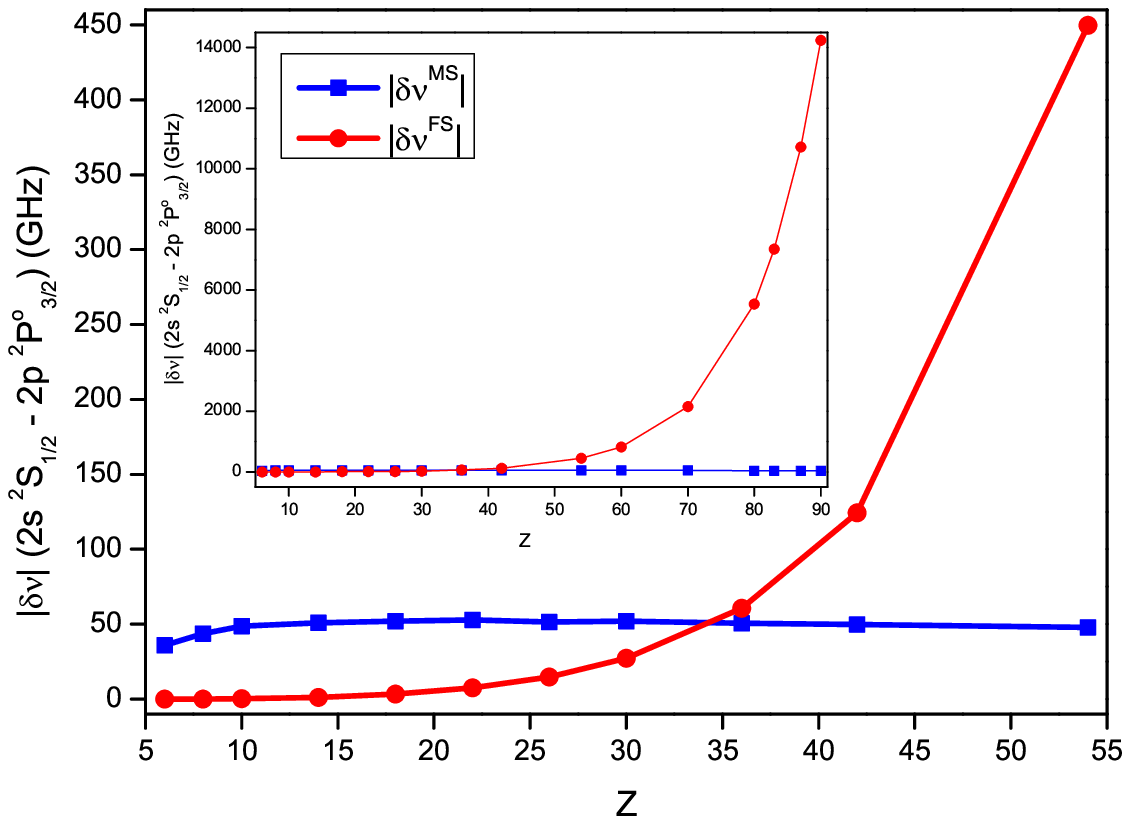}

\caption{\label{fig:2} Absolute value  of the mass shifts $| \delta \nu^{MS}|$  (in blue)  and the  field shifts $\vert \delta \nu^{FS} \vert $ (in red) as functions of the nuclear charge $6 \le Z \le 90$. All values in  GHz. Results are shown for
$2s \;\, ^2S_{1/2} - 2p \;\, ^2P^o_{1/2}$  (upper figure) and
$2s \;\, ^2S_{1/2} - 2p \;\, ^2P^o_{3/2}$   (lower figure), calculated
with correlated wave functions (MCDF+Breit) for the isotope pair $(A, A+1)$.}
\end{figure}

Since the relativistic mass shift  $\Delta \tilde{K}^{RMS}$ and the field shift  $F$ factors are also $Z$-dependent, the mass shifts $\nu^{MS}$ and the field shifts $\nu^{FS}$ can be semi-quantitatively obtained as functions of the atomic number $Z$ with the assistance of Eq. (\ref{AEF}) and (\ref{NREF}). Using the MCDF+Breit results from table \ref{tab:3} and \ref{tab:4}, we illustrate in figure \ref{fig:2} the absolute value of $|\nu^{MS}|$ and $|\nu^{FS}|$ for the $2s \;\, ^2S_{1/2} - 2p \;\, ^2P^o_{1/2}$ and $2s \;\, ^2S_{1/2} - 2p \;\, ^2P^o_{3/2}$ transitions in Li-like ions with $6 \le Z \le 90$ for the isotope pair $(A,A+1)$. The balance between the mass shifts and the field shifts can be easily observed from these two figures. As can be seen, in the region of $6 \le Z \le 35$ the mass shift is somewhat larger than the field shift for the transitions concerned, but they are of the same order of magnitude. Therefore, both of them  must  be taken into account for a relevant analysis of isotope shifts. Over the isoelectronic sequence, the mass shifts can be regarded as constant, but the field shifts increase drastically towards high-$Z$ values and become dominant at the end of the sequence. For instance, the field shift is one order of magnitude larger than the mass shift for Li-like Xe ion. As a result, one can safely neglect the mass shift for $Z \ge 80$ ions.

In addition, it should be emphasized that the mass shift and the field shift cancel out with each other when building the isotope shifts of the $2s \;\, ^2S_{1/2} - 2p \;\, ^2P^o_{1/2}$ and $2s \;\, ^2S_{1/2} - 2p \;\, ^2P^o_{3/2}$ transitions. Therefore, the total isotope shifts approach zero around $Z=34$ (Se$^{31+}$).

\subsection{Isotope Shifts in $^{150,142}$Nd$^{57+}$}
According to Eq. (\ref{MF-eq}), we calculate the mass shift, the field shift and the total isotope shifts for the $2s \;\, ^2S_{1/2} - 2p \;\, ^2P^o_{1/2}$ and $2s \;\, ^2S_{1/2} - 2p \;\, ^2P^o_{3/2}$ transitions in the case of $^{150,142}$Nd$^{57+}$, using the MCDF+Breit mass- and field-shift factors. In this calculation, the atomic masses~\cite{NIST} are used to compute the mass difference between the nucleus $^{150}$Nd and $^{142}$Nd. The nuclear root-mean-square (rms) value $\langle r^2 \rangle^{1/2}$ taken from Ref.~\cite{Angeli:04} are adopted for estimating the corresponding $\delta \langle r^2 \rangle^{150,142} = 1.2909$~fm$^2$. Since Eq. (\ref{MF-eq}) is based on the first-order perturbation theory, the corresponding results are labelled as ``PT''.
The total isotope shifts for these two transitions are obtained by adding the mass shift and the PT field shift contributions. The experimental measurements \cite{Brandau:08} for the isotope shifts are also displayed in this table. Considering that the contributions from the nuclear deformation~\cite{Kozhedub:08}, nuclear polarization~\cite{Brandau:08}, nuclear size effects on the one-loop QED corrections~\cite{Kozhedub:08} and QED recoil corrections~\cite{Brandau:08, Kozhedub:08, Kozhedub:10} are not taken into account in the present work, they are subtracted from the experimental values to comply with an objective comparison. The corresponding  values are reported in the last line of Table \ref{tab:5} and are marked by a ``$\ast$''. It is found from this table that the mass shift values are in perfect agreement with Brandau \etal{}~\cite{Brandau:08} and Kozhedub \etal{}~\cite{Kozhedub:10} results,  and the isotope shifts are in satisfactory consistence with observations~\cite{Brandau:08}.

One should keep in mind however that the $\delta \langle r^2 \rangle^{150,142} $ value of Angeli has been recently revised by Brandau \etal{} from a combination analysis~\cite{Brandau:08}. Adopting this latest value, ie. $\delta \langle r^2 \rangle^{150,142} =$ 1.36(1)(3) fm$^2$ for estimating the PT theoretical field shift leads to 5~\% discrepancies in the isotope shifts. With respect to the accuracy of calculated field shift factors, it is worthwhile to point out that the higher-order nuclear moments ($\delta \langle r^4 \rangle$, $\delta \langle r^6 \rangle$, \ldots, etc.)~\cite{Seltzer:69} are neglected in the PT field shifts. These missing contributions can be retrieved by estimating the FS from the difference of the transition energies calculated using the specific nuclear potential for each isotope. Adopting the rms values of Angeli \cite{Angeli:04} to build the nuclear potential of both isotopes $^{142}$Nd and $^{150}$Nd, we recalculate the field shifts in the MCDF+Breit model. The corresponding results are marked ``variational'' (VA)  in table~\ref{tab:5}. A difference of $\simeq 5$~\% is found between the PT and VA field shifts, which accounts for the higher-order moments correction. Rescaling our variational field shift with the latest mean-square charge radius difference value gives  $\delta E^{150,142} = -41.18$~meV and $-42.25$~meV for the $2s \;\, ^2S_{1/2} - 2p \;\, ^2P^o_{1/2}$ and $2s \;\, ^2S_{1/2} - 2p \;\, ^2P^o_{3/2}$ transitions, respectively, in very nice agreement  with Brandau \etal{}'s measurements, when taking the nuclear deformation, QED recoil and nuclear contributions into account.

\begin{table}
\caption{Comparison of calculated isotope shifts and parameters for the
$2s \;\, ^2S_{1/2} - 2p \;\, ^2P^o_{1/2}$ and $2s \;\, ^2S_{1/2} - 2p \;\, ^2P^o_{3/2}$ transitions of lithium-like
$^{150,142}$Nd$^{57+}$ with experiment and other computations. All shifts are
given in meV. The nuclear rms radii of these two isotopes are taken from Ref. \cite{Angeli:04}.}\label{tab:5}
\begin{center}
\begin{tabular}{c c c c c c c}
\hline
\hline
                                 && \multicolumn{2}{c}{$2s \;\, ^2S_{1/2} - 2p \;\, ^2P^o_{1/2}$}
                                 && \multicolumn{2}{c}{$2s \;\, ^2S_{1/2} - 2p \;\, ^2P^o_{3/2}$}    \\
\cline{3-4}\cline{6-7}           && This work &  Others \cite{Brandau:08}
                                 && This work &  Others \cite{Brandau:08}    \\
\hline
\multicolumn{2}{c}{\textbf{Mass shift}}                        \\
$\delta \langle H_{RMS} \rangle$ && $1.30$    &  $1.30$
                                 && $1.50$    &  $1.50$        \\[0.2cm]
\hline
\multicolumn{2}{c}{\textbf{Field shift}}                       \\
PT                               && $-42.09$  &
                                 && $-43.55$  &                \\
VA                               && $-40.30$  &
                                 && $-41.70$  &                \\[0.2cm]
\hline
\multicolumn{2}{c}{\textbf{Isotope shift}}                     \\
PT                               && $-40.79$  &
                                 && $-42.05$  &                \\
VA                               && $-39.00$  &
                                 && $-40.20$                   \\
                                 && $-41.18^\dagger$  &
                                 && $-42.45^\dagger$  &        \\  [0.2cm]
Exp.~\cite{Brandau:08}           &&           & $-40.2(3)(6)$
                                 &&           & $-42.3(12)(20)$ \\
                                 &&           & $-41.33^{\ast}$
                                 &&           & $-43.4^{\ast}$  \\
\hline
\hline
\end{tabular}
\end{center}
$^\dagger$VA values rescaled using the  $\delta \langle r^2 \rangle^{150,142}$ from \cite{Brandau:08} instead of \cite{Angeli:04}. \\
$^\ast$Obtained by subtracting the nuclear deformation~\cite{Kozhedub:08}, nuclear polarization~\cite{Brandau:08}, QED recoil~\cite{Brandau:08} and nuclear size effects on the one-loop QED corrections~\cite{Kozhedub:08}  from the experimental values of Brandau \etal{}~\cite{Brandau:08}.
\end{table}

\section{Conclusions} \label{sec:conclu}
Based on the MCDF method, we calculate the mass shift  $\Delta \tilde{K}$ and the field shift  $F$ factors for the $2s \;\, ^2S_{1/2} - 2p \;\, ^2P^o_{1/2}$ and $2s \;\, ^2S_{1/2} - 2p \;\, ^2P^o_{3/2}$ transitions in Li-like ions with $4 \le Z \le 90$. The uncertainties of these parameters are controlled within 1\% in the Breit approximation thanks to the rather complete description of electronic correlation. Our results provide the reader with a fast estimate of the expected IS for a given element and pair of isotopes, using formula (\ref{MF-eq}). If the isotope shift is measured for one of the lines, alternatively, the $\Delta \tilde{K}^{RMS}$- and $F$-parameters can be utilized to extract the change in the mean-square charge radii.

Using these results we analyze the competition between the mass shift and the field shift for the $2s  - 2p $ resonance doublet along the isoelectronic sequence. It is found that the mass shift and the field shift possess similar orders of magnitude in the  $Z < 40$ range, so that one should consider both of them for a relevant analysis of isotope shifts, especially for extracting the nuclear mean-square charge radius. Towards the high-$Z$ region, the field shift grows rapidly and becomes dominant at the end of the sequence. Therefore, the mass shift can be safely neglected in the region of $Z > 80$. In addition, the total isotope shifts approach zero around $Z=35$ due to strong cancellation between  the mass  and the field shift contributions.

A detail analysis of the isotope shifts in the case of $^{150,142}$Nd$^{57+}$ shows that higher-order nuclear moments, often neglected in the calculation of the field shift, should be considered for very highly charged ions in order to extract $\delta \langle r^2 \rangle $ values from experiments.

\section*{ACKNOWLEDGMENTS}

J.G. Li and M. Godefroid thank the Communaut\'e fran\c{c}aise of Belgium (Action de Recherche Concert\'ee) and the Belgian National Fund for Scientific Research (FRFC/IISN Convention) for financial support. C. Naz\'e is grateful to the ``Fonds pour la formation \`a la Recherche dans l'Industrie et dans l'Agriculture'' of Belgium for a PhD grant (Boursier F.R.S.-FNRS). P. J\"onsson acknowledges support by the Swedish Research Council.

\end{document}